\documentclass[aps,prb,twocolumn,showpacs]{revtex4}

\usepackage{graphicx}
\usepackage{amsmath}
\usepackage{amssymb}

\begin{document}

\title{Anisotropic microwave conductivity of cuprate superconductors
in the presence of CuO chain induced impurities}

\author{Zhi Wang and Shiping Feng$^{*}$}
\affiliation{Department of Physics, Beijing Normal University,
Beijing 100875, China}

\begin{abstract}
The anisotropy in the microwave conductivity of the ortho-II
YBa$_2$Cu$_3$O$_{6.50}$ is studied within the kinetic energy driven
superconducting mechanism. The ortho-II YBa$_2$Cu$_3$O$_{6.50}$ is
characterized by a periodic alternative of filled and empty
$\hat{b}$-axis CuO chains. By considering the CuO chain induced
extended anisotropy impurity scattering, the main features of the
anisotropy in the microwave conductivity of the ortho-II
YBa$_2$Cu$_3$O$_{6.50}$ are reproduced based on the nodal
approximation of the quasiparticle excitations and scattering
processes, including the intensity and lineshape of the energy and
temperature dependence of the $\hat{a}$-axis and $\hat{b}$-axis
microwave conductivities. Our results also confirm that the
$\hat{b}$-axis CuO chain induced impurity is the main source of the
anisotropy.
\end{abstract}


\maketitle

\section{Introduction}

Impurity plays a crucial role in determining the behavior of many
measurable properties of cuprate superconductors \cite{hirschfeld1},
such as the quasiparticle transport in the superconducting (SC)
state. This follows from a fact that the physical properties of
cuprate superconductors in the SC state are extreme sensitivity to
the impurity effect than the conventional superconductors due to the
finite angular-momentum charge carrier Cooper pairing with the
d-wave symmetry \cite{hirschfeld1}. The single common feature is the
presence of the CuO$_{2}$ plane \cite{shen}, and it seems evident
that the unusual behaviors of cuprate superconductors are dominated
by this CuO$_{2}$ plane \cite{anderson}. Experimentally, over twenty
years measurements of electrodynamic properties at microwave
energies have provided rather detailed information on the
quasiparticle transport of cuprate superconductors
\cite{bonn,sflee,hosseini,turner,harris,bobowski}, where the nodal
quasiparticle spectrum contains most of the features expected for
weak-limit impurity scattering. The early microwave conductivity
measurements \cite{bonn,sflee,hosseini,turner} showed the microwave
conductivity spectrum is CuO$_{2}$ plane isotropic. However, the
recent development of a sufficient number of fixed-energy cavity
perturbation system to map out a coarse microwave conductivity
spectrum \cite{harris,bobowski} allowed to resolve additional
features in the microwave conductivity spectrum. Among these new
achievements is the observation of an anisotropy in the microwave
conductivity spectrum of the highly ordered ortho-II
YBa$_2$Cu$_3$O$_{6.50}$, where the well pronounced difference in
intensity and lineshape in the $\hat{a}$-axis and $\hat{b}$-axis
directions is remarkable.

Although the anisotropy in the microwave conductivity spectrum of
the ortho-II YBa$_2$Cu$_3$O$_{6.50}$ is well-established
experimentally \cite{harris,bobowski}, its full understanding is
still a challenging issue. Theoretically, based on a
phenomenological Bardeen-Cooper-Schrieffer (BCS) formalism with the
d-wave SC gap function, it has been argued in the mean-field level
that the interband transitions produce a strongly anisotropic
feature \cite{Bascones}. However, it has been shown that the
CuO$_{2}$ bilayer may lead to two nearly identical contributions to
both $\hat{a}$-axis and $\hat{b}$-axis microwave conductivities
\cite{harris}, since the splitting between bonding and antibonding
combinations of the planar wavefunctions is expected to be weak
\cite{atkinson}. In our earlier work \cite{wangmicro} based on the
kinetic energy driven SC mechanism \cite{feng1,feng3}, the effect of
the extended impurity scattering potential on the quasiparticle
transport in the non-ortho-II phase of cuprate superconductors has
been discussed within the nodal approximation of the quasiparticle
excitations and scattering processes, and the obtained energy and
temperature dependence of the microwave conductivity are consistent
with the experimental data in the non-ortho-II phase of cuprate
superconductors \cite{bonn,sflee,hosseini,turner}. In the d-wave SC
state of cuprate superconductors, the characteristic feature is the
existence of four nodal points $[\pm\pi/2,\pm\pi/2]$ (in units of
inverse lattice constant) in the Brillouin zone \cite{shen,tsuei},
where the SC gap function vanishes, therefore the CuO$_{2}$ plane
currents are mainly carried by nodal quasiparticles, and the
quasiparticle transport properties of cuprate superconductors in the
SC state are largely governed by the quasiparticle excitations
around the nodes. Since the gap nodes lie close to the diagonals of
the Brillouin zone, these quasiparticle excitations carry both
$\hat{a}$-axis and $\hat{b}$-axis currents, then the microwave
conductivity should be CuO$_{2}$ plane isotropic. However, the
ortho-II YBa$_2$Cu$_3$O$_{6.50}$ is a very special cuprate SC
material, and is characterized by a periodic alternative of filled
and empty $\hat{b}$-axis CuO chains \cite{harris,bobowski}. In
particular, the recent experimental measurements \cite{bobowski}
unambiguously establish that the $\hat{b}$-axis CuO chain induced
impurity is the dominant source of the anisotropy in the microwave
conductivity spectrum of the ortho-II YBa$_2$Cu$_3$O$_{6.50}$. This
experimental result \cite{bobowski} also implies that the
$\hat{b}$-axis CuO chain in the ortho-II YBa$_2$Cu$_3$O$_{6.50}$
induced impurities lead to an anisotropy of the extended impurity
scattering potential in the CuO$_{2}$ planes. In this paper we show
explicitly if the effect of the anisotropy of the extended impurity
scattering potential induced by the $\hat{b}$-axis CuO chain is
considered within the kinetic energy driven SC mechanism, one can
reproduce some main features of the anisotropy in the microwave
conductivity spectrum observed \cite{harris,bobowski} experimentally
on the ortho-II YBa$_2$Cu$_3$O$_{6.50}$.

The rest of this paper is organized as follows. In Sec. II, we
present the basic formalism, where the BSC-like Green's function
under the kinetic energy driven SC mechanism is dressed via the
extended anisotropy impurity scattering. Within this framework, we
calculate explicitly the $\hat{a}$-axis and $\hat{b}$-axis microwave
conductivities based on the nodal approximation of the quasiparticle
excitations and scattering processes. The energy and temperature
dependence of the $\hat{a}$-axis and $\hat{b}$-axis microwave
conductivities of the ortho-II YBa$_2$Cu$_3$O$_{6.50}$ are presented
in Sec. III. Sec. IV is devoted to a summary.

\section{Theoretical Framework}

The basic element of cuprate superconductors is two-dimensional
CuO$_{2}$ planes \cite{shen} as mentioned above. It has been shown
that the essential physics of the doped CuO$_{2}$ plane is properly
accounted by the $t$-$J$ model on a square lattice
\cite{shen,anderson},
\begin{eqnarray}
H=-t\sum_{i\hat{\eta}\sigma}C^{\dagger}_{i\sigma}
C_{i+\hat{\eta}\sigma}+\mu\sum_{i\sigma}C^{\dagger}_{i\sigma}
C_{i\sigma}+J\sum_{i\hat{\eta}}{\bf S}_{i}\cdot{\bf S}_{i+
\hat{\eta}},~~~
\end{eqnarray}
where $\hat{\eta}=\pm\hat{x},\pm \hat{y}$, $C^{\dagger}_{i\sigma}$
($C_{i\sigma}$) is the electron creation (annihilation) operator,
${\bf S}_{i}=(S^{x}_{i}, S^{y}_{i}, S^{z}_{i})$ is spin operator,
and $\mu$ is the chemical potential. This $t$-$J$ model is subject
to an important local constraint $\sum_{\sigma}
C^{\dagger}_{i\sigma}C_{i\sigma}\leq 1$ to avoid the double
occupancy \cite{anderson}, which can be treated properly in
analytical calculations within the charge-spin separation (CSS)
fermion-spin theory \cite{feng2,feng3}, where the constrained
electron operators are decoupled as $C_{i\uparrow}=
h^{\dagger}_{i\uparrow} S^{-}_{i}$ and $C_{i\downarrow}=
h^{\dagger}_{i\downarrow} S^{+}_{i}$, with the spinful fermion
operator $h_{i\sigma}= e^{-i\Phi_{i\sigma}}h_{i}$ describes the
charge degree of freedom together with some effects of spin
configuration rearrangements due to the presence of the doped charge
carrier itself, while the spin operator $S_{i}$ describes the spin
degree of freedom, then the electron local constraint for the single
occupancy is satisfied in analytical calculations. In this CSS
fermion-spin representation, the $t$-$J$ model (1) can be expressed
as,
\begin{eqnarray}
H&=&t\sum_{i\hat{\eta}}(h^{\dagger}_{i+\hat{\eta}\uparrow}h_{i\uparrow}
S^{+}_{i}S^{-}_{i+\hat{\eta}}+h^{\dagger}_{i+\hat{\eta}\downarrow}
h_{i\downarrow}S^{-}_{i}S^{+}_{i+\hat{\eta}})\nonumber\\
&-&\mu\sum_{i\sigma} h^{\dagger}_{i\sigma}h_{i\sigma}+J_{{\rm eff}}
\sum_{i\hat{\eta}} {\bf S}_{i}\cdot {\bf S}_{i+\hat{\eta}},
\end{eqnarray}
with $J_{{\rm eff}}=(1-\delta)^{2}J$, and $\delta=\langle
h^{\dagger}_{i\sigma}h_{i\sigma}\rangle=\langle h^{\dagger}_{i}
h_{i}\rangle$ is the charge carrier doping concentration. For a
understanding of the SC state properties of cuprate superconductors,
the kinetic energy driven SC mechanism has been developed
\cite{feng1,feng3}, where the interaction between charge carriers
and spins from the kinetic energy term in the $t$-$J$ model (2)
induces the charge carrier pairing state with the d-wave symmetry by
exchanging spin excitations, then the electron Cooper pairs
originating from the charge carrier pairing state are due to the
charge-spin recombination, and their condensation reveals the SC
ground-state. In particular, it has been shown that this SC state is
a conventional BCS like with the d-wave symmetry \cite{guo1}, so
that the basic BCS formalism with the d-wave SC gap function is
still valid in quantitatively reproducing all main low energy
features of the SC coherence of quasiparticles, although the pairing
mechanism is driven by the kinetic energy by exchanging spin
excitations. Following our previous discussions \cite{wangmicro},
the electron Green's function in the SC state can be obtained in the
Nambu representation as,
\begin{eqnarray}
\tilde{G}({\bf k},\omega)=Z_{F}{\omega\tau_{0}+\bar{\Delta}_{Z}({\bf
k}) \tau_{1}+\bar{\varepsilon}_{{\bf k}}\tau_{3}\over\omega^{2}-
E_{{\bf k}}^{2}},
\end{eqnarray}
where $\tau_{0}$ is the unit matrix, $\tau_{1}$ and $\tau_{3}$ are
Pauli matrices, other notations are defined as same as in Ref.
\cite{wangmicro}, and have been determined by the self-consistent
calculation \cite{feng1,feng3}.

In the presence of impurities, the unperturbed electron Green's
function (3) is dressed via the impurity scattering
\cite{durst,nunner,duffy},
\begin{eqnarray}
\tilde{G}_{I}({\bf k},\omega)&=&\sum_{\alpha}\tilde{G}_{I\alpha}
\tau_{\alpha}({\bf k},\omega)\nonumber\\
&=&[\tilde{G}({\bf k},\omega)^{-1} -\tilde{\Sigma}({\bf k},
\omega)]^{-1},~~~
\end{eqnarray}
with the self-energy $\tilde{\Sigma}({\bf k},\omega)=\sum_{\alpha}
\Sigma_{\alpha}({\bf k},\omega)\tau_{\alpha}$. It has been shown
that all but the scalar component of the self-energy function can be
neglected or absorbed into $\bar{\Delta}_{Z}({\bf k} )$
\cite{durst,nunner,duffy}. In this case, the dressed electron
Green's function (4) can be explicitly rewritten as,
\begin{widetext}
\begin{eqnarray}
\tilde{G}_{I}({\bf k},\omega)=Z_{F}{[\omega-\Sigma_{0}({\bf k},
\omega)]\tau_{0}+\bar{\Delta}_{Z}({\bf k})\tau_{1}+
[\bar{\varepsilon}_{{\bf k}}+\Sigma_{3}({\bf k},\omega)]\tau_{3}
\over [\omega-\Sigma_{0}({\bf k},\omega)]^{2}
-\bar{\varepsilon}^{2}_{{\bf k}}-\bar{\Delta}^{2}_{Z}({\bf k})},~
\end{eqnarray}
\end{widetext}
where the self-energies $\Sigma_{0}({\bf k},\omega)$ and
$\Sigma_{3}({\bf k},\omega)$ are treated within the framework of the
T-matrix approximation as,
\begin{eqnarray}
\tilde{\Sigma}({\bf k},\omega)=\rho_{i}\tilde{T}_{{\bf k}{\bf k}}
(\omega),
\end{eqnarray}
where $\rho_{i}$ is the impurity concentration, and $\tilde{T}_{{\bf
k} {\bf k}}(\omega)$ is the diagonal element of the T-matrix,
\begin{eqnarray}
\tilde{T}_{{\bf k}{\bf k}'}(\omega)=V_{{\bf k}{\bf k}'}\tau_{3}+
\sum_{{\bf k}''}V_{{\bf k}{\bf k}''}\tau_{3}\tilde{G}_{I}({\bf k}'',
\omega)\tilde{T}_{{\bf k}'' {\bf k}'}(\omega),
\end{eqnarray}
with $V_{{\bf k}{\bf k}'}$ is the impurity scattering potential and
$\tilde{T}_{{\bf k}{\bf k}'}(\omega)=T^{0}_{{\bf k}{\bf k}'}
(\omega)\tau_{0}+T^{3}_{{\bf k}{\bf k}'}(\omega)\tau_{3}$. In our
earlier work without considering the $\hat{b}$-axis CuO chain
induced impurities \cite{wangmicro}, we have discussed the effect of
the extended isotropy impurity scattering potential on the
quasiparticle transport in the non-ortho-II phase of cuprate
superconductors within the nodal approximation of quasiparticle
excitations and scattering processes, where there is no gap to the
quasiparticle excitations at the four nodes, and then the
quasiparticles are generated only around these four nodes. In this
CuO$_{2}$ plane isotropic case, a general scattering potential
$V_{{\bf k}{\bf k}'}$ need only be evaluated in three possible cases
\cite{durst,nunner}: the intranode impurity scattering $V_{{\bf k}
{\bf k}'}=V_{1}$ (${\bf k}$ and ${\bf k}'$ at the same node), the
adjacent-node impurity scattering $V_{{\bf k}{\bf k}'}=V_{2}$ (${\bf
k}$ and ${\bf k}'$ at the adjacent nodes), and the opposite-node
impurity scattering $V_{{\bf k}{\bf k}'} =V_{3}$ (${\bf k}$ and
${\bf k}'$ at the opposite nodes). However, this CuO$_{2}$ plane
isotropic case is broken in the presence of the $\hat{b}$-axis CuO
chain induced impurities in the ortho-II YBa$_2$Cu$_3$O$_{6.50}$,
since the CuO chain induced impurity scattering potential can be
described by an anisotropic potential in the CuO$_{2}$ planes
\cite{graser}. This simply assumes that the screened Coulomb
potential created by the CuO chain induced impurities produces a
footprint sensed by quasiparticles moving in the CuO$_{2}$ planes
\cite{graser}. After incorporating this $\hat{b}$-axis CuO chain
induced anisotropy impurity scattering into the extended isotropy
impurity scattering potential, the total scattering potential
$V_{{\bf k}{\bf k}'}$ need be evaluated in four possible cases as
shown in Fig. 1: the intranode impurity scattering $V_{{\bf k} {\bf
k}'}=V_{1}$ (${\bf k}$ and ${\bf k}'$ at the same node) and the
opposite-node impurity scattering $V_{{\bf k} {\bf k}'}=V_{3}$
(${\bf k}$ and ${\bf k}'$ at the opposite nodes), these two cases
are the same as in the previous discussions in the non-ortho-II
phase of cuprate superconductors \cite{wangmicro}. However, the
adjacent-node impurity scattering along the $\hat{a}$-axis direction
$V_{{\bf k}{\bf k}'}=V_{2}$ (${\bf k}$ and ${\bf k}'$ at the
$\hat{a}$-axis adjacent nodes) is different from that along the
$\hat{b}$-axis direction $V_{{\bf k} {\bf k}'}=V_{4}$ (${\bf k}$,
and ${\bf k}'$ at the $\hat{b}$-axis adjacent nodes) in the present
discussions of the ortho-II YBa$_2$Cu$_3$O$_{6.50}$. In this case,
the impurity scattering potential $V_{{\bf k}{\bf k}'}$ in the
T-matrix is effectively reduced as,
\begin{eqnarray}
V_{{\bf k}{\bf k}'}\rightarrow\underline{V}=\left(
\begin{array}{cccc}
V_{1} & V_{2} & V_{3} & V_{4} \\
V_{2} & V_{1} & V_{4} & V_{3} \\
V_{3} & V_{4} & V_{1} & V_{2} \\
V_{4} & V_{3} & V_{2} & V_{1}
\end{array} \right) \,.
\end{eqnarray}

\begin{figure}[t]
\begin{center}
\leavevmode
\includegraphics[clip=true,width=0.5\columnwidth]{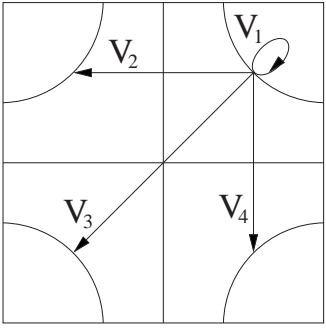}
\caption{The anisotropic impurity scattering within the d-wave case.
$V_{1}$, $V_{2}$, $V_{3}$, and $V_{4}$ are potentials for intranode,
adjacent-node along the $\hat{a}$-axis direction, opposite-node, and
adjacent-node impurity scattering along the $\hat{b}$-axis direction
scattering.} \label{fig:NodalApprox}
\end{center}
\end{figure}

We emphasize that the anisotropy of the impurity scattering
potential (then CuO$_{2}$ plane microwave conductivity) has been
reflected by this important difference between the adjacent-node
impurity scattering strengths $V_{2}$ and $V_{4}$. Now we follow our
previous discussions \cite{wangmicro}, and obtain explicitly the
$\hat{a}$-axis and $\hat{b}$-axis microwave conductivities of the
ortho-II YBa$_{2}$Cu$_{3}$O$_{6+y}$ as,
\begin{widetext}
\begin{subequations}
\begin{eqnarray}
\sigma_a(\omega,T)&=&4e^{2}v^{2}_{f}\int^{\infty}_{-\infty}{d\omega'
\over2\pi}{n_{F}(\omega')-n_{F}(\omega' + \omega) \over\omega'}
[{\rm Re}J_a(\omega'-i\delta,\omega'+\omega+i\delta)-{\rm Re}
J_a(\omega'+i\delta,\omega'+\omega+i\delta)], \\
\sigma_b(\omega,T)&=&4e^{2}v^{2}_{f}\int^{\infty}_{-\infty}{d\omega'
\over2\pi}{n_{F}(\omega')-n_{F}(\omega' + \omega) \over \omega'}
[{\rm Re}J_b(\omega'-i\delta,\omega'+\omega+i\delta)-{\rm Re}
J_b(\omega'+i\delta,\omega'+\omega+i\delta)],
\end{eqnarray}
\end{subequations}
\end{widetext}
where $n_{F}(\omega)$ is the fermion distribution function,
$v_{f}=\sqrt{2}t$ is the electron velocity at the nodal points, and
the kernel function $J_a(\omega',\omega)$ and $J_b(\omega',\omega)$
are given by,
\begin{widetext}
\begin{subequations}
\begin{eqnarray}
J_a(\omega',\omega)={I^{(0)}_{0}+L^a_{1}[I^{(0)}_{0}I_{3}^{(3)}
+I_{0}^{(3)}I_{3}^{(0)}]\over [1-(L^a_{1}I_{0}^{(0)}+L^a_{2}
I^{(0)}_{3})][1-(L^a_{1}I_{3}^{(3)}+L^a_{2}I^{(3)}_{0})]-[L^a_{1}
I_{0}^{(3)}+L^a_{2}I^{(3)}_{3}][L^a_{1}I_{3}^{(0)}
+L^a_{2}I^{(0)}_{0}]},~~~~~\\
J_b(\omega',\omega)={I^{(0)}_{0}+L^b_{1}[I^{(0)}_{0}I_{3}^{(3)}
+I_{0}^{(3)}I_{3}^{(0)}]\over [1-(L^b_{1}I_{0}^{(0)}+L^b_{2}
I^{(0)}_{3})][1-(L^b_{1}I_{3}^{(3)}+L^b_{2}I^{(3)}_{0})]-[L^b_{1}
I_{0}^{(3)}+L^b_{2}I^{(3)}_{3}][L^b_{1}I_{3}^{(0)}
+L^b_{2}I^{(0)}_{0}]},~~~~~
\end{eqnarray}
\end{subequations}
\end{widetext}
where the functions $L^a_1$, $L^a_2$, $L^b_1$, and $L^a_2$ are
expressed as,
\begin{widetext}
\begin{subequations}
\begin{eqnarray}
L^a_1(\omega',\omega)&=&\rho_{i}[T_{11}^0(\omega')
T_{11}^0(\omega'+\omega)+T_{11}^3(\omega')T_{11}^3(\omega'+\omega)
-T_{12}^0(\omega')T_{12}^0(\omega'+\omega)-T_{12}^3(\omega')
T_{12}^3(\omega'+\omega) \nonumber\\
&-&T_{13}^0(\omega')T_{13}^0(\omega'+\omega)-T_{13}^3(\omega')
T_{13}^3(\omega'+\omega)+T_{14}^0(\omega')T_{14}^0(\omega'+\omega)
+T_{14}^3(\omega')T_{14}^3(\omega'+\omega)],
\end{eqnarray}
\begin{eqnarray}
L^a_2(\omega',\omega)&=&\rho_{i}[T_{11}^0(\omega')
T_{11}^3(\omega'+\omega)+T_{11}^3(\omega')T_{11}^0(\omega'+\omega)
-T_{12}^0(\omega')T_{12}^3(\omega'+\omega)-T_{12}^3(\omega')
T_{12}^0(\omega'+\omega)\nonumber\\
&-&T_{13}^0(\omega')T_{13}^3(\omega'+\omega)
-T_{13}^3(\omega')T_{13}^0(\omega'+\omega)+T_{14}^0(\omega')
T_{14}^3(\omega'+\omega)+T_{14}^3(\omega')T_{14}^0(\omega'+\omega)],
\\
L^b_1(\omega',\omega)&=&\rho_{i}[T_{11}^0(\omega')
T_{11}^0(\omega'+\omega)+T_{11}^3(\omega')T_{11}^3(\omega'+i\omega)
+T_{12}^0(\omega')T_{12}^0(\omega'+\omega)+T_{12}^3(\omega')
T_{12}^3(\omega'+\omega)\nonumber\\
&-&T_{13}^0(\omega')T_{13}^0(\omega'+\omega)-T_{13}^3(\omega')
T_{13}^3(\omega'+\omega)-T_{14}^0(\omega')T_{14}^0(\omega'+\omega)-
T_{14}^3(\omega')T_{14}^3(\omega'+\omega)],\\
L^b_2(\omega',\omega)&=&\rho_{i}[T_{11}^0(\omega')
T_{11}^3(\omega'+\omega)+T_{11}^3(\omega')T_{11}^0(\omega'+\omega)
+T_{12}^0(\omega')T_{12}^3(\omega'+\omega)+T_{12}^3(\omega')
T_{12}^0(\omega'+\omega)\nonumber\\
&-&T_{13}^0(\omega')T_{13}^3(\omega'+\omega)-T_{13}^3(\omega')
T_{13}^0(\omega'+\omega)-T_{14}^0(\omega')T_{14}^3(\omega'+\omega)-
T_{14}^3(\omega')T_{14}^0(\omega'+\omega)],
\end{eqnarray}
\end{subequations}
\end{widetext}
and the functions $I^{(0)}_{0}(\omega,\omega')$ and
$I^{(0)}_{3}(\omega,\omega')$ are evaluated in terms of the dressed
Green's function (5) as,
\begin{widetext}
\begin{subequations}
\begin{eqnarray}
I^{(0)}_{0}(\omega,\omega')\tau_{0}&+&I^{(3)}_{0}(\omega,\omega')
\tau_{3} ={1\over N}\sum_{{\bf k}}\tilde{G}_{I}({\bf k},\omega)
\tilde{G}_{I}({\bf k},\omega+\omega'), \\
I^{(0)}_{3}(\omega,\omega')\tau_{0}&+&I^{(3)}_{3}(\omega,\omega')
\tau_{3}={1\over N}\sum_{{\bf k}}\tilde{G}_{I}({\bf k},\omega)
\tilde{\tau}_{3}\tilde{G}_{I}({\bf k},\omega+\omega').
\end{eqnarray}
\end{subequations}
\end{widetext}
It is clearly that if the effect of the $\hat{b}$-axis CuO chain
induced impurities is neglected, i.e., $V_2=V_4$, this leads to
$L_1^a(\omega',\omega)=L_1^b(\omega',\omega)$ and
$L_2^a(\omega',\omega)=L_2^b(\omega',\omega)$, and then the
$\hat{a}$-axis and $\hat{b}$-axis microwave conductivities in Eq.
(9) are reduced to the isotropic one \cite{wangmicro}
$\sigma_a(\omega',\omega)= \sigma_b(\omega', \omega)=
\sigma(\omega',\omega)$.

\section{Energy and temperature dependence of the $\hat{a}$-axis and
$\hat{b}$-axis microwave conductivities for the ortho-II
YBa$_2$Cu$_3$O$_{6.50}$}

In cuprate superconductors, although the values of $J$ and $t$ is
believed to vary somewhat from compound to compound \cite{shen},
however, as a qualitative discussion, the commonly used parameters
in this paper are chosen as $t/J=2.5$, with an reasonably estimative
value of $J\sim 1000$K. We are now ready to discuss the energy and
temperature dependence of the $\hat{a}$-axis and $\hat{b}$-axis
quasiparticle transport of the ortho-II YBa$_2$Cu$_3$O$_{6.50}$ with
the extended anisotropy impurity scattering. We have performed a
calculation for the energy dependence of the $\hat{a}$-axis and
$\hat{b}$-axis microwave conductivities $\sigma_{a}(\omega,T)$ and
$\sigma_{b}(\omega,T)$ in Eq. (9) at low temperatures, and the
results of $\sigma_{a}(\omega,T)$ (top panel) and
$\sigma_{b}(\omega,T)$ (bottom panel) as a function of energy with
temperature $T=0.002J=2$K (solid line), $T=0.004J=4$K (dash-dotted
line) and $T=0.008J=8$K (dashed line) under the slightly strong
impurity scattering potential with $V_{1}=100J$, $V_{2}=90J$,
$V_{3}=30J$, and $V_{4}=30J$ at the impurity concentration
$\rho_{i}=0.00002$ for the doping concentration $\delta=0.15$ are
plotted in Fig. 2 in comparison with the corresponding experimental
results \cite{harris} of the ortho-II YBa$_2$Cu$_3$O$_{6.50}$
(inset). It is clearly that the anisotropy of the energy evolution
of the low temperature microwave conductivity of the ortho-II
YBa$_2$Cu$_3$O$_{6.50}$ is qualitatively reproduced
\cite{harris,bobowski}, where quasiparticle spectral weights differ
by a factor of two between the $\hat{a}$-axis and $\hat{b}$-axis
directions. Moreover, although the cusplike lineshapes are observed
along both $\hat{a}$-axis and $\hat{b}$-axis directions as in the
previous case in the non-ortho-II phase of cuprate superconductors
\cite{wangmicro}, the width of the $\hat{a}$-axis spectrum is
significantly broadened. This broadening is only attributed to
increased quasiparticle scattering arising from the CuO chain
induced impurities. In comparison with our previous results for the
non-ortho-II phase of cuprate superconductors \cite{wangmicro}, the
present results of the anisotropy therefore confirm that the CuO
chain induced impurity is the dominant source of the anisotropy in
the microwave conductivity spectrum of the ortho-II
YBa$_2$Cu$_3$O$_{6.50}$ \cite{bobowski}.

\begin{figure}
\includegraphics[scale=0.5]{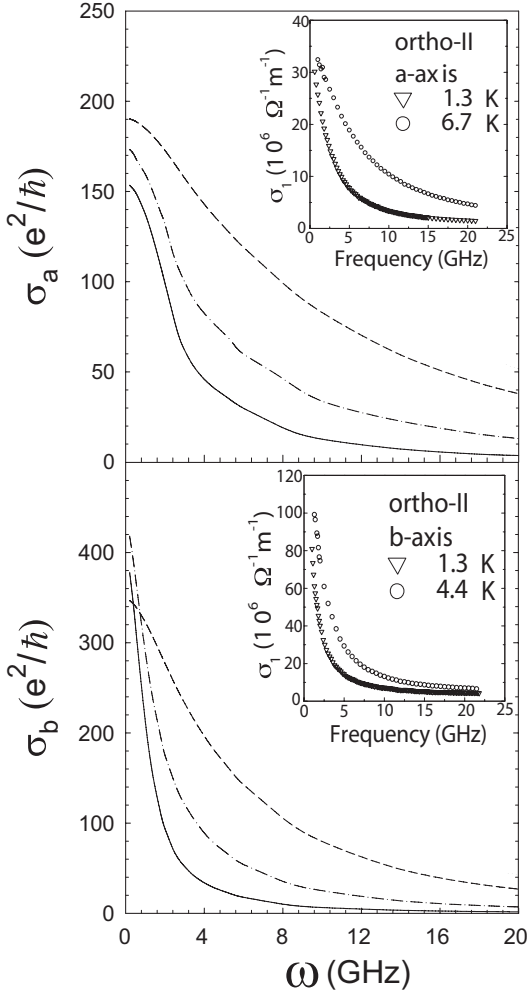}
\caption{The microwave conductivities of the $\hat{a}$-axis (top
panel) and $\hat{b}$-axis (bottom panel) as a function of energy
with $T=0.002J=2$K (solid line), $T=0.004J=4$K (dash-dotted line)
and $T=0.008J =8$K (dashed line) at $\rho_{i}=0.00002$ for $t/J=2.5$
with $V_{1}=100J$, $V_{2}=90J$, $V_{3}=30J$, and $V_{4}=30J$ in
$\delta=0.15$. Inset: the corresponding experimental result of the
ortho-II YBa$_2$Cu$_3$O$_{6.50}$ taken from Ref.
\onlinecite{harris}.}
\end{figure}

\begin{figure}
\includegraphics[scale=0.5]{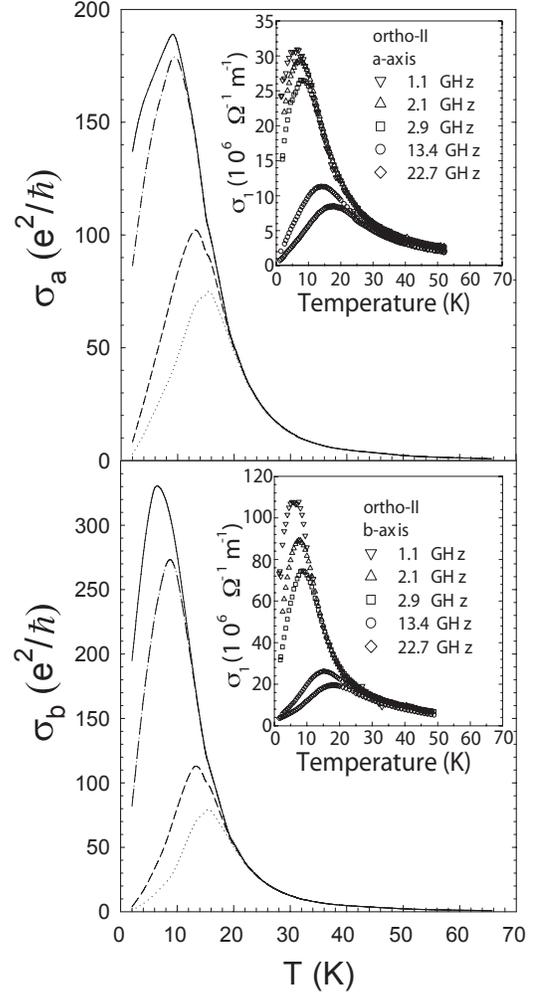}
\caption{The microwave conductivities of the $\hat{a}$-axis (top
panel) and $\hat{b}$-axis (bottom panel) as a function of
temperature with energy $\omega\approx 1.14$GHz (solid line),
$\omega\approx 2.28$GHz (dash-dotted line), $\omega\approx 13.4$GHz
(dashed line), and $\omega\approx 22.8$GHz (dotted line) at
$\rho_{i}=0.00002$ for $t/J=2.5$ with $V_{1}=100J$, $V_{2}=90J$,
$V_{3}=30J$, and $V_{4}=30J$ in $\delta=0.15$. Inset: the
corresponding experimental result of the ortho-II
YBa$_2$Cu$_3$O$_{6.50}$ taken from Ref. \onlinecite{harris}.}
\end{figure}

In the above discussions of the low temperature and low energy case,
the inelastic quasiparticle-quasiparticle scattering process has
been dropped, since it is suppressed at low temperatures and low
energies due to the large SC gap parameter in the quasiparticle
excitation spectrum. For a better understanding of the anisotropy in
the microwave conductivity spectrum, we now turn to discuss the
temperature dependence of the quasiparticle transport of the
ortho-II YBa$_2$Cu$_3$O$_{6.50}$, where $T$ may approach to $T_{c}$
from low temperature side, and therefore the inelastic
quasiparticle-quasiparticle scattering process should be considered
\cite{nunner}. The contribution from this inelastic
quasiparticle-quasiparticle scattering process is increased rapidly
when $T$ approaches to $T_{c}$ from low temperature side, since
there is a small SC gap parameter near $T_{c}$. In particular, it
has been pointed out \cite{walker} that the contribution from the
quasiparticle-quasiparticle scattering process to the transport
lifetime is exponentially suppressed at low temperatures, and
therefore the effect of this inelastic quasiparticle-quasiparticle
scattering can be considered by adding the inverse transport
lifetime \cite{duffy} $\tau^{-1}_{{\rm inel}}(T)$ to the imaginary
part of the self-energy function $\Sigma_{0}(\omega)$ in Eq. (5) as
in our previous discussions \cite{wangmicro}, then the total
self-energy function $\Sigma^{{\rm tot}}_{0}(\omega)$ can be
expressed as \cite{walker,nunner,wangmicro} $\Sigma^{{\rm tot}}_{0}
(\omega)=\Sigma_{0}(\omega)-i[2\tau_{{\rm inel}}(T)]^{-1}$, with
$\tau_{{\rm inel}}(T)$ has been chosen as $[2\tau_{{\rm inel}}
(T)]^{-1}=2\times10^4(T-0.002)^{4}J$. Using this total self-energy
function $\Sigma^{{\rm tot}}_{0} (\omega)$ to replace
$\Sigma_{0}(\omega)$ in Eq. (5), we have performed a calculation for
the temperature dependence of the $\hat{a}$-axis and $\hat{b}$-axis
microwave conductivities $\sigma_{a}(\omega,T)$ and
$\sigma_{b}(\omega,T)$, and the results of $\sigma_{a}(\omega,T)$
and $\sigma_{b}(\omega,T)$ as a function of temperature $T$ with
energy $\omega=0.0000547J\approx 1.14$GHz (solid line),
$\omega=0.0001094J \approx 2.28$GHz (dash-dotted line),
$\omega=0.0006564J\approx 13.4$GHz (dashed line), and
$\omega=0.001094J\approx 22.8$GHz (dotted line) under the slightly
strong impurity scattering potential with $V_{1}=100J$, $V_{2}=90J$,
$V_{3}=30J$ and $V_{4}=30J$ at the impurity concentration
$\rho_{i}=0.00002$ for the doping concentration $\delta=0.15$ are
plotted in Fig. 3. For comparison, the corresponding experimental
results \cite{harris} of the ortho-II YBa$_2$Cu$_3$O$_{6.50}$ are
also plotted in Fig. 3 (inset). Obviously, the overall temperature
dependence is similar to that of the temperature dependence of the
microwave conductivity in the non-ortho-II phase of cuprate
superconductors \cite{wangmicro}, where both $\hat{a}$-axis and
$\hat{b}$-axis temperature dependent microwave conductivities
$\sigma_{a}(\omega,T)$ and $\sigma_{b}(\omega,T)$ increases rapidly
with increasing temperatures to a broad peak, and then falls roughly
linearly. In particular, this broad peak shifts to higher
temperatures as the energy increased. However, the spectral width is
evidently anisotropic, where the spectral width of the
$\hat{a}$-axis microwave conductivity $\sigma_{a}(\omega,T)$ is
larger than the corresponding value of the $\hat{b}$-axis one.
Moreover, the peak position of the $\hat{a}$-axis microwave
conductivity $\sigma_{a}(\omega,T)$ is located at higher temperature
than the corresponding value of the $\hat{b}$-axis one, in
qualitative agreement with the experimental data of the ortho-II
YBa$_2$Cu$_3$O$_{6.50}$ \cite{harris,bobowski}.

In our present theory, the anisotropy of the microwave conductivity
reflects directly from the electron vertex correction due to the
extended anisotropy impurity scattering potential. However, it has
been shown that for all scattering strength the thermal vertex
correction is negligible compared to the electron one \cite{durst},
therefore we expects the heat transport of the ortho-II
YBa$_2$Cu$_3$O$_{6.50}$ to be CuO$_{2}$ plane isotropic. This is a
unique feature of our theory that is different from the other theory
\cite{Bascones} in which the fermi surface is assumed to be altered
and thus the heat transport may be also anisotropic. This should be
verified by further experiments.

\section{Summary}

Within the framework of the kinetic energy driven SC mechanism, we
have studied the anisotropy in the microwave conductivity spectrum
recently observed  \cite{harris,bobowski} in the ortho-II
YBa$_2$Cu$_3$O$_{6.50}$. The ortho-II YBa$_2$Cu$_3$O$_{6.50}$ is
characterized by a periodic alternative of filled and empty
$\hat{b}$-axis CuO chains. The extended anisotropy impurity
scattering potential results when the extended isotropy impurity
scattering potential in the CuO$_{2}$ planes incorporates the
$\hat{b}$-axis CuO chain induced impurity scattering potential.
Based on the nodal approximation of the quasiparticle excitations
and scattering processes, we have calculated the $\hat{a}$-axis and
$\hat{b}$-axis microwave conductivities with the extended anisotropy
impurity scattering, and reproduced the main features
\cite{harris,bobowski} of the anisotropy in the microwave
conductivity spectrum of the ortho-II YBa$_2$Cu$_3$O$_{6.50}$,
including the intensity and lineshape of the energy and temperature
dependence of the $\hat{a}$-axis and $\hat{b}$-axis microwave
conductivities. Our results also confirm that the $\hat{b}$-axis CuO
chain induced impurity is the main source of the anisotropy
\cite{bobowski}.

\acknowledgments

This work was supported by the National Natural Science Foundation
of China under Grant No. 10774015, and the funds from the Ministry
of Science and Technology of China under Grant Nos. 2006CB601002 and
2006CB921300.


\begin{references}
\bibitem [*] {add} To whom correspondence should be addressed.
\end{references}
\end{document}